# Journal topic citation potential and between-field comparisons: The topic normalized impact factor

Pablo Dorta-González [a], María Isabel Dorta-González [b], Dolores Rosa Santos-Peñate [a], Rafael Suárez-Vega [a]

[a] Instituto de Turismo y Desarrollo Económico Sostenible Tides, Universidad de Las Palmas de Gran Canaria, Spain; [b] Departamento de Estadística, Investigación Operativa y Computación, Universidad de La Laguna, Spain.

**ABSTRACT**

The journal impact factor is not comparable among fields of science and social science because of systematic differences in publication and citation behaviour across disciplines. In this work, a source normalization of the journal impact factor is proposed. We use the aggregate impact factor of the citing journals as a measure of the citation potential in the journal topic, and we employ this citation potential in the normalization of the journal impact factor to make it comparable between scientific fields. An empirical application comparing some impact indicators with our topic normalized impact factor in a set of 224 journals from four different fields shows that our normalization, using the citation potential in the journal topic, reduces the between-group variance with respect to the within-group variance in a higher proportion than the rest of indicators analysed. The effect of journal self-citations over the normalization process is also studied.

## 1. Introduction

This work is related to journal metrics and citation-based indicators for the assessment of scientific scholar journals from a general bibliometric perspective. For decades, the *journal impact factor* (JIF) has been an accepted indicator in ranking journals. However, there are increasing arguments against the fairness of using the JIF as the sole ranking criteria (Waltman & Van Eck, 2013).

The *2-year impact factor* published by Thomson Reuters in the *Journal Citation Reports* (JCR) is defined as the average number of citations to each journal in a current year with respect to 'citable items' published in that journal during the two preceding years (Garfield, 1972). Nevertheless, it has been criticized due to arbitrary decisions in its construction. The definition of 'citable items' including letters together with the peer reviewed papers (research articles, proceedings papers, and reviews), the focus on the two preceding years, the incomparability between fields, etc., have been discussed in the literature (Bensman, 2007; Moed et al., 2012) and have given many possible modifications and improvements (Althouse et al., 2009; Bornmann & Daniel, 2008). In response, Thomson Reuters has incorporated the *5-year impact factor,* the *eigenfactor score*, and the *article influence score* (Bergstrom, 2007) to the JCR journals. All these indicators consider a 5-year citation window and are useful for comparing journals in the same subject category. However, subject categories may overlap and are sometimes problematic. Moreover, although in many cases the 5-year impact factor is greater than the 2-year impact factor, both indicators lead statistically to the same ranking (Leydesdorff, 2009; Rousseau, 2009). Alternative indicators, considering at the same time production and impact, are the *central area indices* (Dorta-González & Dorta-González, 2010, 2011; Egghe, 2013).

Nevertheless, all the previous impact indicators do not solve the problem when comparing journals from different fields of science. Different scientific fields have different citation practices and citation-based bibliometric indicators need to be normalized for such differences in order to allow for journal comparisons. This problem of field-specific differences in citation impact indicators comes from institutional research evaluation (Leydesdorff & Bornmann, 2011; Van Raan et al., 2010). For example, research institutes often have among their missions the objective of integrating interdisciplinary bodies of knowledge which are generally populated by scholars with different disciplinary backgrounds (Leydesdorff & Rafols, 2011; Wagner et al., 2011).

There are statistical patterns which are field-specific and allow for the normalization of the JIF. Garfield (1979) proposes the term ‘citation potential’ for systematic differences among fields of science, based on the average number of references. For example, in the biomedical fields long reference lists with more than fifty items are common, but in mathematics short lists with less than twenty references are the standard (Dorta-González & Dorta-González, 2013a). These differences are a consequence of the citation cultures and can produce significant differences in the JIF, since the probability of being cited is affected. In this sense, the average number of references is the variable that has most frequently been used in the literature to justify the differences between fields of science, as well as the most employed in source-normalization (Leydesdorff & Bornmann, 2011; Moed, 2010; Zitt & Small, 2008). However, the variables that to a greater degree explain the variance in the impact factor do not include the average number of references (Dorta-González & Dorta-González, 2013a) and therefore it is necessary to consider other sources of variance in the normalization process, such as the ratio of references to journals included in the JCR, the field growth, the ratio of JCR references to the target window, and the proportion of cited to citing items. Given these large differences in citation practices, the development of bibliometric indicators that allow for between-field comparisons is clearly a critical issue (Waltman & Van Eck, 2013).

Traditionally, normalization for field differences has usually been done based on a field classification system. In said approach, each publication belongs to one or more fields and the citation impact of a publication is calculated relative to the other publications in the same field. Most efforts to classify journals in terms of fields of science have focused on correlations between citation patterns (Leydesdorff, 2006; Rosvall & Bergstrom, 2008). An example of a field classification system is the *JCR subject category list* (Pudovkin & Garfield, 2002; Rafols & Leydesdorff, 2009). For these subject categories, Egghe & Rousseau (2002) propose the *aggregate impact factor* in a similar way as the JIF, taking all journals in a category as one meta-journal. However, the position of individual journals of merging specialties remains difficult to determine with precision and some journals are assigned to more than one category. In this sense, Dorta-González & Dorta-González (2013a) propose the *categories normalized impact factor* considering all the indexing categories of each journal.

Nevertheless, the precise delineation between fields of science and the next-lower level specialties has until now remained an unsolved problem in bibliometrics because these delineations are fuzzy at any moment in time and develop dynamically over time. Therefore, classifying a dynamic system in terms of fixed categories can lead to error because the classification system is defined historically while the dynamics of science is evolutionary (Leydesdorff, 2012, p.359).

Recently, the idea of source normalization was introduced, which offers an alternative approach to normalizing field differences. In this approach, normalization is achieved by looking at the referencing behaviour of citing journals. Journal performance is a complex multi-dimensional concept difficult to be fully captured in one single metric (Moed et al., 2012, p. 368). In this sense many indices, such as the *fractionally counted impact factor* (Leydesdorff & Bornmann, 2011; Zitt & Small, 2008), dividing each citation by the number of references, and the *2-year maximum journal impact factor* (Dorta-González & Dorta-González, 2013b), considering the 2-year citation time window of maximum impact instead of the previous 2-year time window, have been proposed. Other indicators for the Scopus database, with a 3-year citation time window and a different definition of citable items, are the *source normalized impact per paper SNIP* (Moed, 2010), dividing each citation by the median number of references, and the *scimago journal ranking SJR* (González-Pereira et al., 2011), considering the prestige of the citing journals.

However, all these metrics do not include any great degree of normalization in relation to the specific topic of each journal. The topic normalization is necessary because different scientific topics have different citation practices. Therefore, citation-based bibliometric indicators need to be normalized for such differences between topics in order to allow for between-topic comparisons of the citation impact. In this sense, we use the aggregate impact factor of the citing journals as a measure of the citation potential in the journal topic, and we employ this citation potential in the normalization of the journal impact factor to make it comparable between scientific fields. In order to test this new impact indicator, an empirical application with more than two hundred journals belonging to four different fields is presented. As the main conclusion, we obtain that our *topic normalized impact factor* reduces the between-group variance in relation to the within-group variance in a higher proportion than the rest of indicators analysed, as well as not being influenced by journal self-citations.

## 2. The normalization of the impact factor using the citation potential in the journal topic

The editorial policy of a journal determines its explicit topic. However, the implicit topic can be determined by its scientific impact. In this sense, we can define the topic of the citation impact of a journal, hereafter journal topic, through all the citing journals. For example, if a journal *j* is cited by journals in *n* different fields, then the journal topic can be characterized by all these *n* fields in a proportional form to the number of citations to journal *j*.

We define the citation potential in the topic of journal *j* in a year *y* as the weighted average of the impact factors of all citing journals to *j* in the year *y* with respect to the previous two years. This average is weighted by the number of citations to *j*, excluding self-citations of *j* to *j*.

However, why does this citation potential characterize the journal topic? Given two journals with the same impact factor, the journal of the topic with less citation potential is more influential. This is because the probability of being cited is affected by the systematic differences in the citation cultures among topics.

The idea of normalizing the impact factor of a journal through all citing journals does not intend to assess each citation by the influence or prestige of the citing journal, but characterizes the journal topic in terms of its citation potential and uses it in the normalization process.

In this section we formulate a source normalization, considering the citation potential in the journal topic. We divide the JIF by the citation potential in the journal topic. Thus, if the JIF is higher than the citation potential in its topic then this ratio will be higher than the JIF, whereas if the JIF is smaller than the citation potential in its topic then this ratio will be smaller than the JIF.

In order to facilitate the reading of the formulation in the rest of this section, Table 1 shows the notation with its explanation.

[Table 1 about here]

### *2.1 The journal impact factor*

A journal impact indicator is a measure of the number of times that items published in a census period, cite items published during an earlier target window. The impact factor reported by Thomson Reuters has a one year census period and uses the two previous years as the target window.

As an average, the impact factor is based on two elements: the numerator, which is the number of citations in the current year to any items published in a journal in the previous two years, and the denominator, which is the number of ‘citable items’ (articles, proceedings papers, reviews, and letters) published in the same previous two years (Garfield, 1972). Journal items include ‘citable items’ but also editorials, news, corrections, retractions, and other items.

Let $NPub_y^j$ be the number of publications (citable items) in journal *j* in year *y*. Let $NCit_{y,y-t}^j$ be the number of times in year *y* that the year *y-t* volumes of journal *j* are cited by journals in the database, *t=1, 2*. Then, the journal impact factor of *j* in year *y* is:

$$JIF_y^j = \frac{NCit_{y,y-1}^j + NCit_{y,y-2}^j}{NPub_{y-1}^j + NPub_{y-2}^j}. \qquad (1)$$

*2.2 The citation potential of a database*

As a reference measure in the normalization process we propose the citation potential of the database. This measure will be later used in the normalization weighting factor.

Let *J* be the set of all journals in a specific database (e.g. Web of Science, Scopus, etc.) Denoting $NPub_y^J = \sum_{j \in J} NPub_y^j$ and $NCit_{y,y-t}^J = \sum_{j \in J} NCit_{y,y-t}^j$, the citation potential in *J* is the ratio between the citations in year *y* to any journal of database *J* in years *y-1* and *y-2*, and the number of citable items published in years *y-1* and *y-2*, that is,

$$CP_y^J = \frac{\sum_{j \in J}\left(NCit_{y,y-1}^j + NCit_{y,y-2}^j\right)}{\sum_{j \in J}\left(NPub_{y-1}^j + NPub_{y-2}^j\right)} = \frac{NCit_{y,y-1}^J + NCit_{y,y-2}^J}{NPub_{y-1}^J + NPub_{y-2}^J}. \qquad (2)$$

This citation potential can also be expressed as a weighted average impact factor considering weights proportional to the number of citable items in the target years. Let

$$v_y^j = \frac{NPub_{y-1}^j + NPub_{y-2}^j}{NPub_{y-1}^J + NPub_{y-2}^J} \qquad (3)$$

be the weight of journal $j$ in the database $J$ in the target window of year $y$. Note that $\sum_{j\in J} v_y^j = 1$.

Then, from Equations 1 to 3,

$$
\begin{aligned}
CP_y^J &= \frac{\sum_{j\in J}\left(NCit_{y,y-1}^j + NCit_{y,y-2}^j\right)}{NPub_{y-1}^J + NPub_{y-2}^J} = \sum_{j\in J} \frac{NCit_{y,y-1}^j + NCit_{y,y-2}^j}{NPub_{y-1}^J + NPub_{y-2}^J} \\
&= \sum_{j\in J}\left( \frac{NPub_{y-1}^j + NPub_{y-2}^j}{NPub_{y-1}^J + NPub_{y-2}^J} \times JIF_y^j \right) = \sum_{j\in J}\left( v_y^j \times JIF_y^j \right).
\end{aligned} \tag{4}
$$

This formulation allows us to easily obtain the citation potential of the JCR database, which is 2.822 in year 2011 (Dorta-González & Dorta-González, 2013a). It also allows us to calculate, in a similar way, the citation potential in any set of journals (as discussed below).

### *2.3 The citation potential in the journal topic*

Later, a journal topic normalization of the impact factor will be proposed. This normalization is achieved considering the aggregate impact factor in the topic of each journal, which characterizes its citation potential. The citation potential in the topic of a journal $j$ is proposed as a weighted average of the impact factors of all citing journals $i$, excluding self-citations of journal $j$, weighted by the number of citations from $i$ to $j$.

In a more formal way, we define the *topic* of a journal $j \in J$ as the set of all journals $i \in J$ that in the current year $y$ cite the previous 2-years issues *y-1* and *y-2* of journal $j$, excluding journal $j$ self-citations. In this topic the weight of each journal $i$ is proportional to the number of citations from $i$ to $j$.

In this definition, in a similar way as in the impact factor, we exclusively consider citations in the census year $y$ to the target window of years *y-1* and *y-2* as the representation of the topic at the research front. We have proposed a formulation excluding journal self-citation because in some cases the percentage of journal self-citation is so high that it could lead to a normalized impact factor close to the classical JIF. However, the effect of journal self-citation in the normalization process is also studied in the empirical application.

Let $T_j$ be the topic of journal $j$, that is, the meta-journal of all citing journals to journal $j$ excluding journal $j$. Let $NCit_{y,y-t}^{ij}$ be the number of times in year $y$ that the year *y-t* volumes of journal $j$ are cited by journal $i$ in the database $J$, $t=1, 2$. Therefore, the weight of journal $i$ in the topic of journal $j$ in year $y$ is:

$$w_y^{ij} = \frac{NCit_{y,y-1}^{ij} + NCit_{y,y-2}^{ij}}{\sum_{k \in T_j} \left( NCit_{y,y-1}^{kj} + NCit_{y,y-2}^{kj} \right)}. \qquad (5)$$

Note that $\sum_{i \in T_j} w_y^{ij} = 1, \ \forall j \in J.$

Therefore, in a similar way as in Equation 4, the formulation of the citation potential in the topic of journal $j$ (i.e. the aggregate impact factor of meta-journal $T_j$) as a weighted average impact factor is:

$$CP_y^{T_j} = \sum_{i \in T_j} \left( w_y^{ij} \times JIF_y^i \right). \qquad (6)$$

This aggregate impact factor is a measure of the citation potential in the topic of journal $j$. Later, it will be used in the normalization of the indicator.

Consider the example in Figure 1. Let $j$ be a journal with JIF = 2.000 and the citing journals (excluding $j$) indicated in Figure 1. The citation potential in the topic of journal $j$ is 0.5×1.000 + 0.3×2.500 + 0.15×0.800 + 0.05×1.400 = 1.440. The journal impact factor (2.000) is 39% greater than the citation potential in the topic (2.000 / 1.440 = 1.39) and, therefore, in the comparison with other journals the JIF should be proportionally increased in a way that will be illustrated below.

[Figure 1 about here]

### *2.4 The Topic Normalized Impact Factor*

We propose a normalized citation indicator that compares 'actual' impact factor with 'expected' impact factor, based on the citation potential of its topic, i.e., the weighted average impact factor of all citing journals.

The ratio $CP_y^J \,/\, CP_y^{T_j}$ is the normalized score in the topic of journal $j$. If $CP_y^J = CP_y^{T_j}$ then this score is one. A score higher than one shows that the citation potential in the journal topic is below the citation potential in the database, while a score lower than one shows that the citation potential in the journal topic is above the citation potential in the database.

Therefore, we define the *Topic Normalized Impact Factor* of journal $j$ in year $y$ as:

$$TNIF_y^j = \frac{CP_y^J}{CP_y^{T_j}} \times JIF_y^j.$$

In the case where $CP_y^{T_j} = 0$ we consider that $TNIF_y^j = 0$. Notice that if $CP_y^{T_j} > CP_y^J$ then the score is lower than one and therefore it reduces the impact factor of journal *j*. Conversely, if $CP_y^{T_j} < CP_y^J$ then the score is higher than one and therefore it increases the impact factor of journal *j*.

In the example of Figure 1, considering that $CP_y^J = 1.800$ then the normalized score of journal *j* is $CP_y^J \,/\, CP_y^{T_j} = 1.800 / 1.440 = 1.25$ and the $TNIF_y^j = 1.25 \times 2.000 = 2.500$. This amount is greater than the JIF because the citation potential of the database is greater than the citation potential in the topic of the journal.

## 3. Methods and Materials

We used six impact indicators: 2-year journal impact factor (2-JIF), 5-year journal impact factor (5-JIF), eigenfactor score (ES), fractionally counted impact factor (FCIF), topic normalized impact factor (TNIF), and TNIF including self-citation (Self-cite).

We designed a cluster sample. Cluster sampling is a two-stage sampling design in which, firstly, one single cluster is randomly selected from a set of clusters and, secondly, all observations in the selected cluster are included in the sample (Bornmann & Mutz, 2013). Four fields (journal categories), each one from a different cluster obtained by Dorta-González & Dorta-González (2013a), were considered. This was motivated in order to obtain journals with systematic differences in publication and citation behaviour. A total of 224 journals were considered in this empirical application. The journal categories and the number of journals in each category are: Astronomy & Astrophysics (56); Biology (85); Engineering, Aerospace (27); and History & Philosophy of Science (56).

The bibliometric data was obtained from the online version of the 2011 *Journal Citation Reports* (JCR) during the first week of May 2013. The JCR database (reported by Thomson Reuters, USA) is available at the www.webofknowledge.com website.

The fractionally counted impact factor was obtained from Leydesdorff & Bornmann (2011) because this indicator is calculated on a different database than the JCR.

## 4. Results and Discussion

In the empirical application we studied which impact indicator produces a closer data distribution among scientific fields in relation to its centrality and variability measures. We used six impact indicators: 2-year journal impact factor (2-JIF), 5-year journal impact factor (5-JIF), eigenfactor score (ES), fractionally counted impact factor (FCIF), topic normalized impact factor (TNIF), and TNIF with self-citation (Self-cite).

Table 2 shows the impact indicators for a set of 224 journals from four fields. Two of the fields have higher impact factors (Astronomy & Astrophysics and Biology), while the other two have lower impact factors (Aerospace Engineering and History & Philosophy of Science). Notice the ampleness in the variation interval for each indicator. For example, the range is 26.452 for the 2-JIF and 29.657 for the 5-JIF. In the case of the TNIF, the range of variation is 31.237, while this range is 13.572 when considering journal self-citation. This means that removing the influence of journal self-citation produces an increase in the variability of the scores, and therefore, the discrimination ability of the indicator increases.

[Table 2 about here]

The general pattern in Table 2 is a 5-JIF higher than the 2-JIF. Moreover, in those fields with lower impact factors (Aerospace Engineering and History & Philosophy of Science) there is a higher increase in the TNIF in relation to the JIF. This effect reduces the differences between fields in the case of the TNIF.

Notice the ampleness in the variation interval for the citation potential in the journal topics. The score varies from 1.736 to 6.049 in Astronomy & Astrophysics, from 0.345 to 5.993 in Biology, from 0 to 2.952 in Aerospace Engineering, and from 0 to 6.777 in History & Philosophy of Science. Note the citation potentials in the journal topics are very different from one another even within the same field. This means that the journal topic is one possible explanatory factor in the variance of the impact indicators. This variance may also reflect differences in quality between the journals or the publication of certain document types (e.g. reviews) in some journals. Moreover, the difference in the score, with and without self-citation, is very relevant in many cases and above one in ten journals. Note the case of P NATL A SCI INDIA where this difference is 2.320, and J BIOL EDUC where this is 1.955, for example.

The citation potential of the journal topic has an inverse effect over the topic normalized impact factor. That is, the lower the citation potential of the journal topic, the greater the increment in the topic normalized impact factor and vice versa. With respect to the self-citation effect, in some cases the self-citation increases the citation potential of the journal topic, thereby reducing the TNIF, but in other cases it reduces the citation potential of the journal topic, thereby increasing the TNIF.

Tables 3 and 4 provide the Pearson correlations and the Spearman rank correlations for all pairs of indicators, both for journal categories and aggregate data. The fact that a perfect Spearman correlation results when the two indexes are related by any monotonic function, can be contrasted with the Pearson correlation, which only gives a perfect value when the two indexes are related by a linear function. In this sense, the Spearman correlation is less sensitive than the Pearson correlation to strong outliers that are in the tails of both distributions. This is because Spearman coefficient limits the outlier to the value of its rank.

[Tables 3 and 4 about here]

We consider the three typical levels of confidence: 99%, 95%, and 90% (significance levels of 0.01, 0.05, and 0.10). In 55 out of 75 possible cases in Table 3 (Pearson correlations) the confidence level is above 99%, and in 4 cases it is above 90%. In the other 16 cases the confidence level is below 90%. However, in 66 out of 75 possible cases in Table 4 (Spearman correlations) the confidence level is above 99%, and in 6 cases it is above 90%. Only in 3 of the 75 possible cases the confidence level is below 90%.

The correlation coefficients are interpreted according to the guidelines of Cohen (1992). The square of the correlation coefficient (coefficient of determination) is the proportion of variance in either of the two variables which may be predicted by (or attributed to) the variance of the other, using a straight-line relationship. For example, when $r = 0.85$, $r^2 = 0.72$, the 72% of the variance in the dependent variable is attributable to the independent variable. Cohen (1988: 77-81) states as a guiding criterion in the behavioral sciences: small effect size $r = 0.10$, medium effect size $r = 0.30$, and large effect size $r = 0.50$. According to this criterion, in Table 3, there is large effect size in 43 out of 75 cases, medium effect size in 18 out of 75 cases, and small effect size in 14 out of 75 cases.

The general pattern that can be observed in the correlations reported in Table 3 is that 2-JIF and 5-JIF are very strongly correlated, with all of the Pearson correlations above 0.94. For the aggregate data this correlation is 0.99 and the 2-JIF can explain more than 98% of the variance in the 5-JIF ($0.99^2 = 0.98$). The two TNIFs, including and excluding self-citation, have in most cases correlations much larger than 0.85. In the aggregate data, each indicator explains more than 72% of the variance in the other ($0.85^2 = 0.72$).

The correlations between indicators are higher in those journal categories in which the impact factors are high (Astronomy & Astrophysics and Biology), with 15 out of 30 coefficients larger than 0.85, and lower in those in which the impact factors are low (Aerospace Engineering and History & Philosophy of Science), with 7 (Pearson) and 4 (Spearman) out of 30 coefficients larger than 0.85.

Central-tendency and variability measures for the fields are showed in Table 5. All the indicators have skewed distributions, with many journals having low values and only a small number of journals with high values. This is the reason why in these skewed distributions the medians are well below means. Notice the high differences between categories in medians, means, and standard deviations.

[Table 5 about here]

The fields considered are very different in relation to the citation behavior and some of them are penalized by the JIF. Note that the central-tendency measures of the JIF in Astronomy & Astrophysics and Biology are very much higher than those in Aerospace Engineering and History & Philosophy of Science; in general, more than three times higher. However, the central-tendency measures of the TNIF are closer in all the fields considered. Furthermore, removing the influence of journal self-citation produces an increment in the variability of the scores.

Finally, we tested if the journal topic normalization reduces the between-group variance in relation to the within-group variance. Table 6 shows the central-tendency measures for the aggregate data and the between-group variances. The between-group or explained variance is the variability that is produced by the independent variable, i.e., the group differences. The within-group or error variance is the variability that is not produced by the independent variable. Note that the journal topic normalization produces the greatest percentage reduction of the variance (94.4%). Moreover, removing the influence of journal self-citation produces an increment in the within-group variance of the scores and therefore a better indicator discrimination ability.

[Table 6 about here]

## 5. Conclusions

Different scientific fields have different citation practices, and citation-based bibliometric indicators need to be normalized for such differences between fields in order to allow for between-field comparisons of citation indicators. In this paper, we provide a source normalization approach based on the journal topic and we compare it with some popular impact indicators.

An empirical application, with more than two hundred journals from four different fields, shows that our journal topic normalization reduces the between-group variance in relation to the within-group variance more than the rest of the indicators analyzed in this paper.

The fields considered are very different in relation to the citation behavior. For this reason, the JIF in Astronomy & Astrophysics and Biology are very much higher than the JIF in Aerospace Engineering and History & Philosophy of Science. However, the TNIFs are very close in all the fields considered. We propose removing the influence of journal self-citation because it produces an increment in the variability of the scores, whereby providing a better indicator discrimination ability.

Finally, it is necessary to be cautious when comparing journal impact indicators from different fields. In this sense, our index has behaved well in a great number of journals from very different fields.

The topic of journal *j* in year *y*

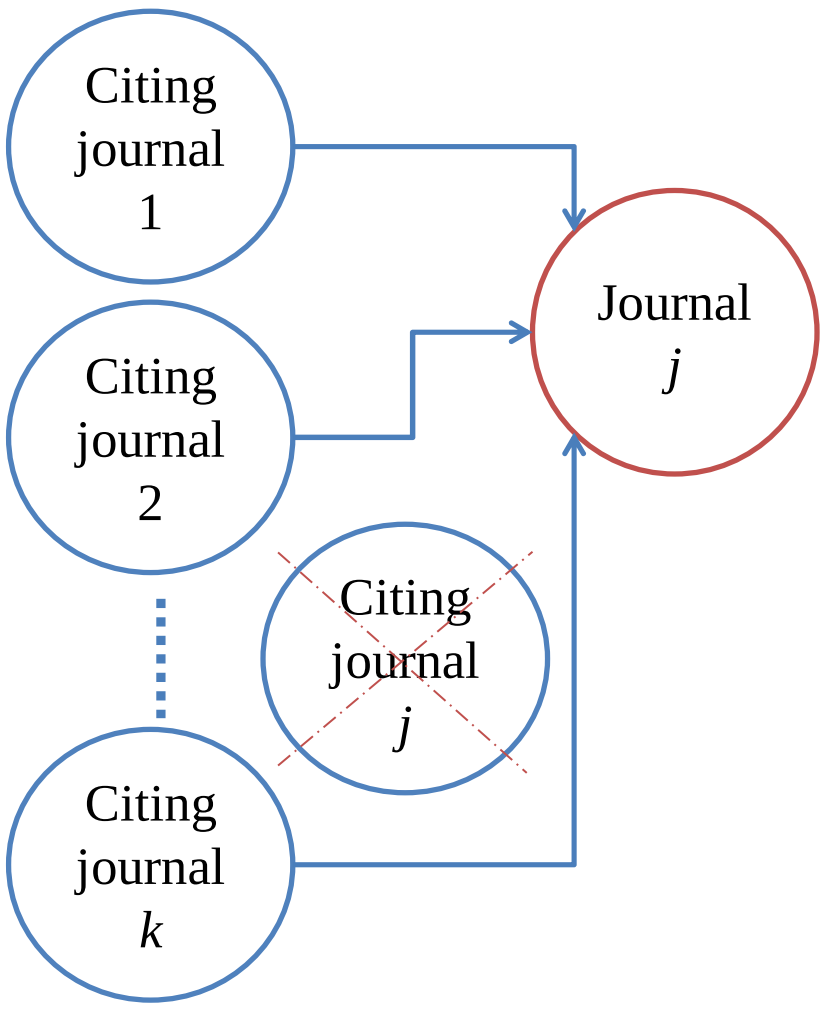

Citing journals in *y* with a weight of total citations in *y* to *y*-1 and *y*-2.

The citation potential of the journal topic

Let *j* be a journal with JIF = 2.000, and consider the following data about the citing journals (excluding *j*):

| Citing journal | Total citations | Weight | JIF |
|---|---|---|---|
| 1 | 100 | 100 / 200 = 0.5 | 1.000 |
| 2 | 60 | 60 / 200 = 0.3 | 2.500 |
| 3 | 30 | 30 / 200 = 0.15 | 0.800 |
| 4 | 10 | 10 / 200 = 0.05 | 1.400 |
| Total | 200 | 1 | |

Citation potential in the topic of journal *j*:
$0.5\times1+0.3\times2.5+0.15\times0.8+0.05\times1.4 = 1.440$.

Then, 2.000 / 1.440 = 1.39 and the JIF (2.000) is 39% greater than the citation potential in the journal topic (1.440).

Figure 1: One example of journal topic and citation potential

Table 1: Notation

| *Notation* | *Explanation* |
|---|---|
| $J$ | Journals in the database |
| $j \in J$ | Journal in evaluation |
| $T_j$ | Topic of journal *j* (meta-journal of all citing journals to *j*, excluding *j*) |
| $i \in T_j$ | Journals in the topic of journal *j* |
| $y$ | Current year (census period) |
| $y-1,\ y-2$ | Citation time window (target window) |
| $NPub_y^j$ | Number of publications (citable items) in journal *j* in year *y* |
| $NPub_y^J = \sum_{j \in J} NPub_y^j$ | Number of publications (citable items) in database *J* |
| $v_y^j = \dfrac{NPub_{y-1}^j + NPub_{y-2}^j}{NPub_{y-1}^J + NPub_{y-2}^J}$ | Weight of journal *j* in database *J* in the target window of year *y* |
| $NCit_{y,y-t}^{ij}$ | Number of times in year *y* that the year *y-t* volumes of journal *j* are cited by the journal *i*, *t=1,2* |
| $w_y^{ij} = \dfrac{NCit_{y,y-1}^{ij} + NCit_{y,y-2}^{ij}}{\sum\limits_{k \in T_j} \left( NCit_{y,y-1}^{kj} + NCit_{y,y-2}^{kj} \right)}$ | Weight of journal *i* in the topic of *j* in year *y* |
| $NCit_{y,y-t}^{j} = \sum_{i \in J} NCit_{y,y-t}^{ij}$ | Number of times in year *y* that the year *y-t* volumes of journal *j* are cited by journals in the database, *t=1,2* |
| $JIF_y^j = \dfrac{NCit_{y,y-1}^{j} + NCit_{y,y-2}^{j}}{NPub_{y-1}^j + NPub_{y-2}^j}$ | Journal impact factor of *j* in year *y* |
| $CP_y^J = \sum_{j \in J} \left( v_y^j \times JIF_y^j \right)$ | Citation potential of database *J* in year *y* (aggregate impact factor in *J*) |
| $CP_y^{T_j} = \sum_{i \in T_j} \left( w_y^{ij} \times JIF_y^i \right)$ | Citation potential of topic *j* in year *y* (aggregate impact factor in topic *j*) |
| $TNIF_y^j = \dfrac{CP_y^J}{CP_y^{T_j}} \times JIF_y^j$ | Topic normalized impact factor of journal j in year y |

Table 2: Impact indicators for a set of 224 journals from four fields

| *Abbreviated journal title* | *JCR Category* | *2-JIF* | *5-JIF* | *ES* | *FCIF* | *Citation Potential* | | *Topic Normalized* | |
|---|---|---|---|---|---|---|---|---|---|
| | | | | | | *Self-cite* | $CP^{Tj}$ | *Self-cite* | $TNIF^{j}$ |
| ACTA ASTRONAUT | EA | 0.614 | 0.619 | 0.00541 | 0.05420 | 0.540 | 0.501 | 3.209 | 3.458 |
| ACTA ASTRONOM | A&A | 1.680 | 2.200 | 0.00227 | 0.22789 | 4.102 | 4.773 | 1.156 | 0.993 |
| ACTA BIOL HUNG | B | 0.593 | 0.625 | 0.00071 | 0.12339 | 1.393 | 1.451 | 1.201 | 1.153 |
| ADV EXP MED BIOL | B | 1.093 | 1.374 | 0.03161 | -- | 0.965 | 0.964 | 3.196 | 3.200 |
| ADV SPACE RES | A&A | 1.178 | 1.066 | 0.01963 | -- | 1.794 | 1.947 | 1.853 | 1.707 |
| AEROBIOLOGIA | B | 1.515 | 1.669 | 0.00110 | -- | 2.101 | 2.308 | 2.035 | 1.852 |
| AERONAUT J | EA | 0.482 | 0.501 | 0.00158 | 0.06042 | 0.884 | 0.929 | 1.539 | 1.464 |
| AEROSP SCI TECHNOL | EA | 0.983 | 0.934 | 0.00293 | 0.13795 | 0.734 | 0.706 | 3.779 | 3.929 |
| AEROSPACE AM | EA | 0.048 | 0.036 | 0.00026 | 0.00668 | 1.481 | 1.481 | 0.091 | 0.091 |
| AGR HIST | H&PS | 0.312 | 0.231 | 0.00017 | -- | 0.176 | 0.137 | 5.003 | 6.427 |
| AGR HUM VALUES | H&PS | 1.540 | 1.717 | 0.00213 | 0.03927 | 1.095 | 1.057 | 3.969 | 4.112 |
| AIAA J | EA | 1.057 | 1.277 | 0.01749 | 0.18514 | 1.190 | 1.226 | 2.507 | 2.433 |
| AIRCR ENG AEROSP TEC | EA | 0.195 | 0.281 | 0.00053 | 0.00188 | 0.881 | 0.881 | 0.625 | 0.625 |
| AM BIOL TEACH | B | 0.133 | 0.199 | 0.00041 | 0.01314 | 0.345 | 0.345 | 1.088 | 1.088 |
| AM J BIOETHICS | H&PS | 4.083 | 3.581 | 0.00482 | 0.32004 | 1.331 | 1.079 | 8.657 | 10.679 |
| AM J HUM BIOL | B | 2.267 | 2.211 | 0.00628 | 0.22760 | 1.444 | 1.342 | 4.430 | 4.767 |
| AMBIX | H&PS | 0.444 | -- | 0.00015 | -- | 0.378 | 0.222 | 3.315 | 5.644 |
| ANN GEOPHYS-GERMANY | A&A | 1.842 | 1.757 | 0.01989 | 0.36827 | 2.443 | 2.546 | 2.128 | 2.042 |
| ANN HUM BIOL | B | 1.975 | 1.789 | 0.00280 | 0.12583 | 1.388 | 1.320 | 4.015 | 4.222 |
| ANN SCI | H&PS | 0.417 | 0.451 | 0.00051 | 0.08108 | 0.575 | 0.594 | 2.047 | 1.981 |
| ANNU REV ASTRON ASTR | A&A | 26.452 | 29.657 | 0.02108 | 2.20106 | 5.500 | 4.944 | 13.572 | 15.099 |
| ANNU REV EARTH PL SC | A&A | 7.227 | 8.850 | 0.01065 | 1.08572 | 3.138 | 3.038 | 6.499 | 6.713 |
| ARCH BIOL SCI | B | 0.360 | -- | 0.00069 | -- | 0.575 | 0.767 | 1.767 | 1.325 |
| ARCH HIST EXACT SCI | H&PS | 0.184 | 0.218 | 0.00019 | 0.06667 | 0.122 | 0.000 | 4.256 | 0.000 |
| ASIA LIFE SCI | B | 0.239 | 0.215 | 0.00008 | -- | 0.446 | 0.524 | 1.512 | 1.287 |
| ASTROBIOLOGY | A&A | 2.150 | 2.806 | 0.00608 | 0.39406 | 2.331 | 2.392 | 2.603 | 2.536 |
| ASTROBIOLOGY | B | 2.150 | 2.806 | 0.00608 | 0.39406 | 2.331 | 2.392 | 2.603 | 2.536 |
| ASTRON ASTROPHYS | A&A | 4.587 | 3.979 | 0.25425 | 0.43125 | 4.723 | 4.789 | 2.741 | 2.703 |
| ASTRON ASTROPHYS REV | A&A | 11.526 | 14.108 | 0.00548 | 1.07239 | 5.614 | 5.586 | 5.794 | 5.823 |
| ASTRON GEOPHYS | A&A | 0.607 | 0.403 | 0.00046 | 0.11694 | 3.114 | 3.197 | 0.550 | 0.536 |
| ASTRON J | A&A | 4.035 | 4.317 | 0.07981 | 0.46921 | 5.020 | 5.210 | 2.268 | 2.186 |
| ASTRON LETT+ | A&A | 0.988 | 0.865 | 0.00258 | 0.19316 | 2.593 | 3.169 | 1.075 | 0.880 |
| ASTRON NACHR | A&A | 1.012 | 0.862 | 0.00675 | 0.13215 | 4.079 | 4.313 | 0.700 | 0.662 |
| ASTRON REP+ | A&A | 0.725 | 0.671 | 0.00187 | 0.17826 | 1.892 | 2.729 | 1.081 | 0.750 |
| ASTROPART PHYS | A&A | 3.216 | 2.783 | 0.01063 | 0.54674 | 4.098 | 4.239 | 2.215 | 2.141 |
| ASTROPHYS BULL | A&A | 0.843 | -- | 0.00060 | -- | 1.854 | 2.830 | 1.283 | 0.841 |
| ASTROPHYS J | A&A | 6.024 | 5.102 | 0.42962 | 0.53774 | 5.217 | 4.769 | 3.259 | 3.565 |
| ASTROPHYS J LETT | A&A | 5.526 | -- | 0.15733 | -- | 5.195 | 5.162 | 3.002 | 3.021 |
| ASTROPHYS J SUPPL S | A&A | 13.456 | 11.438 | 0.07640 | 1.34721 | 5.011 | 4.564 | 7.578 | 8.320 |
| ASTROPHYS SPACE SCI | A&A | 1.686 | 1.344 | 0.01240 | 0.19411 | 2.393 | 2.737 | 1.988 | 1.738 |
| ASTROPHYSICS+ | A&A | 0.467 | 0.409 | 0.00058 | 0.03317 | 2.465 | 4.079 | 0.535 | 0.323 |
| B ASTRON SOC INDIA | A&A | 2.722 | -- | 0.00069 | -- | 4.169 | 4.307 | 1.843 | 1.783 |
| B HIST MED | H&PS | 0.514 | 0.913 | 0.00109 | 0.11131 | 0.188 | 0.155 | 7.715 | 9.358 |
| B MATH BIOL | B | 1.847 | 2.002 | 0.00812 | 0.27554 | 1.623 | 1.605 | 3.211 | 3.247 |
| B STOR SCI MAT | H&PS | 0.000 | -- | 0.00001 | -- | 0.000 | 0.000 | 0.000 | 0.000 |
| BALT ASTRON | A&A | 0.444 | 0.575 | 0.00100 | 0.06481 | 3.356 | 4.909 | 0.373 | 0.255 |
| BER WISSGESCH | H&PS | 0.289 | -- | 0.00016 | -- | 1.098 | 1.401 | 0.743 | 0.582 |
| BIOCELL | B | 0.630 | 0.710 | 0.00057 | 0.06176 | 2.097 | 2.097 | 0.848 | 0.848 |
| BIOELECTROCHEMISTRY | B | 3.759 | 3.238 | 0.00644 | 0.21842 | 2.584 | 2.541 | 4.105 | 4.175 |
| BIOELECTROMAGNETICS | B | 1.842 | 2.165 | 0.00330 | 0.25775 | 1.422 | 1.251 | 3.656 | 4.155 |
| BIOESSAYS | B | 4.954 | 4.754 | 0.02410 | 0.50921 | 1.658 | 1.606 | 8.432 | 8.705 |
| BIOL BULL+ | B | 0.200 | 0.247 | 0.00055 | 0.01550 | 0.470 | 0.680 | 1.201 | 0.830 |
| BIOL BULL-US | B | 1.698 | 2.197 | 0.00414 | 0.32143 | 1.486 | 1.477 | 3.225 | 3.244 |
| BIOL DIRECT | B | 4.017 | 3.860 | 0.00690 | -- | 3.383 | 3.287 | 3.351 | 3.449 |
| BIOL LETTERS | B | 3.762 | 4.049 | 0.02992 | -- | 2.362 | 2.327 | 4.495 | 4.562 |
| BIOL PHILOS | H&PS | 1.203 | 1.360 | 0.00184 | 0.17451 | 1.873 | 1.960 | 1.813 | 1.732 |
| BIOL RES | B | 1.029 | 1.269 | 0.00173 | 0.16880 | 1.472 | 1.521 | 1.973 | 1.909 |
| BIOL REV | B | 9.067 | 11.790 | 0.01402 | 1.39561 | 2.033 | 1.962 | 12.586 | 13.041 |
| BIOL RHYTHM RES | B | 0.440 | 0.593 | 0.00062 | 0.09468 | 1.689 | 2.469 | 0.735 | 0.503 |
| BIOLOGIA | B | 0.557 | 0.630 | 0.00256 | 0.06022 | 0.641 | 0.659 | 2.452 | 2.385 |
| BIOMETRICS | B | 1.827 | 2.249 | 0.02046 | 0.42128 | 0.982 | 1.037 | 5.250 | 4.972 |
| BIOMETRIKA | B | 1.912 | 2.575 | 0.01880 | 0.26280 | 1.055 | 0.997 | 5.114 | 5.412 |
| BIOSCI J | B | 0.215 | -- | 0.00047 | -- | 0.466 | 0.526 | 1.302 | 1.153 |
| BIOSCI TRENDS | B | 0.968 | 0.811 | 0.00070 | -- | 5.508 | 5.508 | 0.496 | 0.496 |
| BIOSCIENCE | B | 4.621 | 6.223 | 0.01816 | 0.59420 | 1.829 | 1.684 | 7.130 | 7.744 |
| BIOSEMIOTICS-NETH | H&PS | 0.444 | 0.439 | 0.00006 | -- | 1.761 | 2.838 | 0.712 | 0.441 |
| BIOSYSTEMS | B | 1.784 | 1.497 | 0.00591 | 0.20750 | 1.635 | 1.611 | 3.079 | 3.125 |
| BMC BIOL | B | 5.750 | 5.841 | 0.01672 | -- | 2.769 | 2.725 | 5.860 | 5.955 |
| BRAZ ARCH BIOL TECHN | B | 0.551 | 0.638 | 0.00220 | 0.05967 | 0.446 | 0.406 | 3.486 | 3.830 |
| BRAZ J BIOL | B | 0.688 | -- | 0.00290 | -- | 0.628 | 0.617 | 3.092 | 3.147 |
| BRAZ J MED BIOL RES | B | 1.129 | 1.381 | 0.00620 | 0.12949 | 0.694 | 0.682 | 4.591 | 4.672 |
| BRIT J PHILOS SCI | H&PS | 1.097 | 1.364 | 0.00180 | 0.32361 | 0.619 | 0.502 | 5.001 | 6.167 |
| CELEST MECH DYN ASTR | A&A | 1.457 | 1.280 | 0.00290 | 0.20297 | 1.826 | 2.117 | 2.252 | 1.942 |
| CENT EUR J BIOL | B | 1.000 | 1.020 | 0.00126 | -- | 1.397 | 1.457 | 2.020 | 1.937 |
| CHINESE J AERONAUT | EA | 0.406 | -- | 0.00086 | -- | 0.498 | 0.531 | 2.301 | 2.158 |
| CHRONOBIOL INT | B | 4.028 | 3.233 | 0.00616 | 0.19980 | 2.314 | 1.041 | 4.912 | 10.919 |
| CLASSICAL QUANT GRAV | A&A | 3.320 | 2.706 | 0.04893 | 0.36447 | 3.846 | 3.960 | 2.436 | 2.366 |
| COMPUT BIOL CHEM | B | 1.551 | 1.525 | 0.00234 | 0.19171 | 2.118 | 2.150 | 2.067 | 2.036 |
| COMPUT BIOL MED | B | 1.089 | 1.302 | 0.00430 | 0.15093 | 1.025 | 1.014 | 2.998 | 3.031 |
| CONFIGURATIONS | H&PS | 0.182 | 0.321 | 0.00026 | -- | 0.000 | 0.000 | 0.000 | 0.000 |
| CONTRIB ASTRON OBS S | A&A | 0.152 | -- | 0.00024 | -- | 3.750 | 4.650 | 0.114 | 0.092 |
| COSMIC RES+ | A&A | 0.387 | 0.367 | 0.00077 | 0.03642 | 1.471 | 1.920 | 0.742 | 0.569 |
| COSMIC RES+ | EA | 0.387 | 0.367 | 0.00077 | 0.03642 | 1.471 | 1.920 | 0.742 | 0.569 |
| CR BIOL | B | 1.533 | 1.826 | 0.00516 | 0.14757 | 1.621 | 1.625 | 2.669 | 2.662 |
| CR PHYS | A&A | 1.360 | 1.401 | 0.00525 | 0.20083 | 1.867 | 1.867 | 2.056 | 2.056 |
| CRYOBIOLOGY | B | 2.062 | 2.199 | 0.00441 | 0.27563 | 1.583 | 1.494 | 3.676 | 3.895 |
| CRYOLETTERS | B | 1.245 | 1.326 | 0.00100 | 0.15272 | 1.434 | 1.526 | 2.450 | 2.302 |
| CRYPTOLOGIA | H&PS | 0.109 | 0.126 | 0.00015 | -- | 0.077 | 0.069 | 3.995 | 4.458 |
| DYNAMIS | H&PS | 0.143 | 0.265 | 0.00032 | -- | 1.221 | 1.761 | 0.331 | 0.229 |
| EARTH MOON PLANETS | A&A | 0.667 | 0.763 | 0.00204 | 0.14357 | 3.887 | 3.929 | 0.484 | 0.479 |
| EARTH SCI HIST | H&PS | 0.167 | -- | 0.00012 | -- | 0.635 | 1.103 | 0.742 | 0.427 |
| ELECTROMAGN BIOL MED | B | 1.148 | 1.109 | 0.00055 | 0.09607 | 1.631 | 1.743 | 1.986 | 1.859 |
| ENDEAVOUR | H&PS | 0.226 | 0.235 | 0.00030 | 0.04423 | 0.113 | 0.000 | 5.644 | 0.000 |
| ENG STUD | H&PS | 1.048 | 1.048 | 0.00011 | -- | 0.458 | 0.183 | 6.457 | 16.161 |
| EPISTEMOLOGIA | H&PS | 0.077 | -- | 0.00000 | -- | 0.039 | 0.000 | 5.572 | 0.000 |
| ESA BULL-EUR SPACE | EA | 1.163 | 1.511 | 0.00167 | -- | 1.724 | 1.724 | 1.904 | 1.904 |
| EUR J SCI THEOL | H&PS | 0.600 | -- | 0.00006 | -- | 0.721 | 0.851 | 2.348 | 1.990 |
| EUR PHYS J H | H&PS | 1.182 | 1.182 | 0.00014 | -- | 5.486 | 6.777 | 0.608 | 0.492 |
| EXCLI J | B | 1.061 | -- | 0.00023 | -- | 3.620 | 3.739 | 0.827 | 0.801 |
| EXP ASTRON | A&A | 1.818 | 1.950 | 0.00155 | -- | 2.432 | 2.584 | 2.110 | 1.985 |
| FASEB J | B | 5.712 | 6.340 | 0.08876 | 0.69213 | 1.213 | 1.122 | 13.289 | 14.367 |
| FOLIA BIOL-KRAKOW | B | 0.657 | 0.673 | 0.00045 | 0.15163 | 1.104 | 1.282 | 1.679 | 1.446 |
| FOLIA BIOL-PRAGUE | B | 1.151 | 1.183 | 0.00090 | 0.13578 | 2.392 | 2.474 | 1.358 | 1.313 |
| FOUND SCI | H&PS | 0.810 | -- | 0.00026 | -- | 1.103 | 2.226 | 2.072 | 1.027 |
| GEN RELAT GRAVIT | A&A | 2.069 | 2.061 | 0.01002 | 0.29447 | 3.163 | 3.245 | 1.846 | 1.799 |
| GEOBIOLOGY | B | 4.111 | 3.669 | 0.00518 | -- | 2.974 | 2.881 | 3.901 | 4.027 |
| GEOPHYS ASTRO FLUID | A&A | 1.000 | 1.146 | 0.00212 | 0.41417 | 3.985 | 4.386 | 0.708 | 0.643 |
| GRAVIT COSMOL-RUSSIA | A&A | 0.460 | -- | 0.00076 | -- | 2.777 | 3.340 | 0.467 | 0.389 |
| HER RUSS ACAD SCI+ | H&PS | 0.252 | 0.338 | 0.00050 | -- | 0.607 | 0.820 | 1.172 | 0.867 |
| HIST HUM SCI | H&PS | 0.621 | 0.563 | 0.00099 | 0.05227 | 0.851 | 0.862 | 2.059 | 2.033 |
| HIST MATH | H&PS | 0.355 | 0.408 | 0.00044 | 0.04301 | 0.596 | 0.796 | 1.681 | 1.259 |
| HIST PHIL LIFE SCI | H&PS | 0.324 | 0.553 | 0.00025 | -- | 1.171 | 1.594 | 0.781 | 0.574 |
| HIST PHILOS LOGIC | H&PS | 0.235 | 0.230 | 0.00008 | 0.08250 | 0.258 | 0.266 | 2.570 | 2.493 |
| HIST REC AUST SCI | H&PS | 0.400 | -- | 0.00009 | -- | 0.485 | 0.485 | 2.327 | 2.327 |
| HIST SCI | H&PS | 0.667 | 0.699 | 0.00077 | 0.15152 | 0.360 | 0.292 | 5.229 | 6.446 |
| HIST STUD NAT SCI | H&PS | 0.440 | 0.643 | 0.00024 | -- | 3.328 | 3.328 | 0.373 | 0.373 |
| HUM BIOL | B | 1.312 | 1.005 | 0.00145 | 0.20210 | 1.597 | 1.636 | 2.318 | 2.263 |
| HYLE | H&PS | 0.500 | 0.310 | 0.00009 | -- | 0.776 | 0.776 | 1.818 | 1.818 |
| ICARUS | A&A | 3.385 | 3.218 | 0.04792 | 0.63786 | 4.181 | 4.618 | 2.285 | 2.069 |
| IEEE AERO EL SYS MAG | EA | 0.297 | 0.337 | 0.00115 | 0.04121 | 0.572 | 0.572 | 1.465 | 1.465 |
| IEEE ANN HIST COMPUT | H&PS | 0.378 | 0.522 | 0.00046 | -- | 1.597 | 2.359 | 0.668 | 0.452 |
| IEEE T AERO ELEC SYS | EA | 1.095 | 1.680 | 0.00751 | 0.20374 | 0.742 | 0.692 | 4.165 | 4.465 |
| INDIAN J EXP BIOL | B | 1.295 | 1.099 | 0.00276 | -- | 0.637 | 0.580 | 5.737 | 6.301 |
| INT J AEROACOUST | EA | 0.943 | -- | 0.00102 | -- | 1.350 | 1.420 | 1.971 | 1.874 |
| INT J ASTROBIOL | A&A | 1.723 | -- | 0.00140 | -- | 3.204 | 3.765 | 1.518 | 1.291 |
| INT J ASTROBIOL | B | 1.723 | -- | 0.00140 | -- | 3.204 | 3.765 | 1.518 | 1.291 |
| INT J MOD PHYS D | A&A | 1.183 | 1.333 | 0.00920 | 0.32595 | 3.013 | 3.102 | 1.108 | 1.076 |
| INT J RADIAT BIOL | B | 2.275 | 2.139 | 0.00565 | 0.25964 | 1.453 | 1.353 | 4.418 | 4.745 |
| INT J SATELL COMM N | EA | 1.645 | 0.924 | 0.00102 | 0.12637 | 2.462 | 2.952 | 1.886 | 1.573 |
| INT J TURBO JET ENG | EA | 0.025 | 0.135 | 0.00008 | 0.01264 | 0.025 | 0.000 | 2.822 | 0.000 |
| ISIS | H&PS | 0.779 | 1.065 | 0.00204 | 0.20833 | 0.323 | 0.242 | 6.806 | 9.084 |
| J AEROS COMP INF COM | EA | 0.281 | -- | 0.00029 | -- | 0.595 | 0.635 | 1.333 | 1.249 |
| J AEROSPACE ENG | EA | 0.697 | 0.924 | 0.00132 | 0.12851 | 1.223 | 1.309 | 1.608 | 1.503 |
| J AGR BIOL ENVIR ST | B | 1.210 | 1.208 | 0.00176 | 0.39844 | 1.569 | 1.626 | 2.176 | 2.100 |
| J AGR ENVIRON ETHIC | H&PS | 1.109 | 1.242 | 0.00099 | 0.04511 | 0.915 | 0.886 | 3.420 | 3.532 |
| J AIRCRAFT | EA | 0.538 | 0.654 | 0.00649 | 0.13768 | 0.909 | 1.088 | 1.670 | 1.395 |
| J AM HELICOPTER SOC | EA | 0.549 | 0.663 | 0.00080 | 0.16508 | 0.717 | 0.886 | 2.161 | 1.749 |
| J ASTRONAUT SCI | EA | 0.286 | 0.546 | 0.00084 | 0.51923 | 0.681 | 0.681 | 1.185 | 1.185 |
| J ASTROPHYS ASTRON | A&A | 0.400 | 0.477 | 0.00061 | 0.19086 | 2.524 | 2.524 | 0.447 | 0.447 |
| J BIOL EDUC | B | 0.391 | 0.600 | 0.00030 | -- | 2.671 | 4.626 | 0.413 | 0.239 |
| J BIOL RES-THESSALON | B | 0.619 | 0.573 | 0.00031 | -- | 1.230 | 1.371 | 1.420 | 1.274 |
| J BIOL RHYTHM | B | 2.934 | 3.114 | 0.00497 | 0.43625 | 3.015 | 3.025 | 2.746 | 2.737 |
| J BIOL SYST | B | 0.570 | 0.694 | 0.00069 | 0.08423 | 1.291 | 1.608 | 1.246 | 1.000 |
| J BIOSCIENCES | B | 1.648 | 2.218 | 0.00521 | 0.17278 | 0.988 | 0.942 | 4.707 | 4.937 |
| J COSMOL ASTROPART P | A&A | 5.723 | 5.107 | 0.05669 | 0.33380 | 4.512 | 4.102 | 3.579 | 3.937 |
| J ETHNOBIOL | B | 0.576 | -- | 0.00035 | -- | 2.996 | 2.996 | 0.543 | 0.543 |
| J EXP BIOL | B | 2.996 | 3.301 | 0.04616 | 0.49252 | 1.611 | 1.371 | 5.248 | 6.167 |
| J GUID CONTROL DYNAM | EA | 0.941 | 1.159 | 0.00792 | 0.23918 | 0.665 | 0.504 | 3.993 | 5.269 |
| J HIST ASTRON | H&PS | 0.238 | 0.179 | 0.00026 | 0.04321 | 0.915 | 1.457 | 0.734 | 0.461 |
| J HIST BIOL | B | 0.628 | 0.542 | 0.00080 | 0.12593 | 3.752 | 3.917 | 0.472 | 0.452 |
| J HIST BIOL | H&PS | 0.628 | 0.542 | 0.00080 | 0.12593 | 3.752 | 3.917 | 0.472 | 0.452 |
| J HIST MED ALL SCI | H&PS | 0.714 | 0.781 | 0.00067 | 0.07292 | 0.375 | 0.307 | 5.373 | 6.563 |
| J HIST NEUROSCI | H&PS | 0.425 | 0.538 | 0.00032 | -- | 2.441 | 3.113 | 0.491 | 0.385 |
| J KOREAN ASTRON SOC | A&A | 0.615 | -- | 0.00028 | -- | 3.216 | 4.354 | 0.540 | 0.399 |
| J MATH BIOL | B | 2.963 | 2.480 | 0.00784 | 0.29717 | 1.261 | 1.196 | 6.631 | 6.991 |
| J PROPUL POWER | EA | 0.761 | 1.003 | 0.00624 | 0.23808 | 1.216 | 1.425 | 1.766 | 1.507 |
| J RADIAT RES | B | 1.683 | 1.794 | 0.00362 | 0.22551 | 1.352 | 1.308 | 3.513 | 3.631 |
| J SPACECR TECHNOL | EA | 0.000 | -- | 0.00003 | -- | 0.000 | 0.000 | 0.000 | 0.000 |
| J SPACECRAFT ROCKETS | EA | 0.557 | 0.685 | 0.00428 | 0.13376 | 0.816 | 0.879 | 1.926 | 1.788 |
| J THEOR BIOL | B | 2.208 | 2.415 | 0.03209 | 0.33615 | 1.389 | 1.269 | 4.486 | 4.910 |
| J THERM BIOL | B | 1.373 | 1.320 | 0.00241 | 0.21808 | 1.501 | 1.540 | 2.581 | 2.516 |
| KINEMAT PHYS CELEST+ | A&A | 0.361 | -- | 0.00031 | -- | 1.942 | 2.779 | 0.525 | 0.367 |
| LIFE SCI J | B | 0.073 | -- | 0.00015 | -- | 0.845 | 0.845 | 0.244 | 0.244 |
| LIVING REV SOL PHYS | A&A | 12.500 | -- | 0.00278 | -- | 5.339 | 5.019 | 6.607 | 7.028 |
| MATH BIOSCI | B | 1.540 | 1.683 | 0.00671 | 0.20534 | 0.933 | 0.854 | 4.658 | 5.089 |
| MATH MED BIOL | B | 1.818 | 1.604 | 0.00118 | 0.23683 | 1.970 | 1.974 | 2.604 | 2.599 |
| MED HIST | H&PS | 0.535 | 0.545 | 0.00055 | 0.07246 | 0.619 | 0.631 | 2.439 | 2.393 |
| MICROGRAVITY SCI TEC | EA | 0.591 | 0.526 | 0.00126 | 0.31337 | 1.022 | 1.259 | 1.632 | 1.325 |
| MICROSC RES TECHNIQ | B | 1.792 | 1.873 | 0.00662 | 0.27657 | 1.029 | 0.973 | 4.915 | 5.197 |
| MON NOT R ASTRON SOC | A&A | 4.900 | 4.585 | 0.24884 | 0.44058 | 5.030 | 5.118 | 2.749 | 2.702 |
| NEW ASTRON | A&A | 1.411 | 1.327 | 0.00396 | 0.31242 | 3.832 | 4.010 | 1.039 | 0.993 |
| NEW ASTRON REV | A&A | 1.321 | 0.874 | 0.00406 | 0.07966 | 4.269 | 4.269 | 0.873 | 0.873 |
| NEXUS NETW J | H&PS | 0.070 | -- | 0.00006 | -- | 0.053 | 0.000 | 3.727 | 0.000 |
| NOTES REC ROY SOC | H&PS | 0.163 | 0.234 | 0.00027 | 0.07600 | 0.258 | 0.306 | 1.783 | 1.503 |
| NUNCIUS | H&PS | 0.038 | 0.068 | 0.00020 | -- | 0.513 | 0.513 | 0.209 | 0.209 |
| OBSERVATORY | A&A | 0.481 | 0.320 | 0.00013 | 0.23101 | 2.476 | 4.186 | 0.548 | 0.324 |
| ORIGINS LIFE EVOL B | B | 2.660 | 2.081 | 0.00222 | 0.20001 | 5.722 | 5.993 | 1.312 | 1.253 |
| OSIRIS | H&PS | 0.292 | 0.554 | 0.00033 | 0.13636 | 0.390 | 0.390 | 2.113 | 2.113 |
| P BIOL SOC WASH | B | 0.292 | 0.402 | 0.00048 | 0.08753 | 0.533 | 0.593 | 1.546 | 1.390 |
| P I MECH ENG G-J AER | EA | 0.488 | 0.579 | 0.00199 | 0.04850 | 0.665 | 0.752 | 2.071 | 1.831 |
| P NATL A SCI INDIA B | B | 0.019 | -- | 0.00005 | -- | 2.339 | 4.659 | 0.023 | 0.012 |
| P ROY SOC B-BIOL SCI | B | 5.415 | 5.670 | 0.09614 | 0.02516 | 2.297 | 2.129 | 6.653 | 7.178 |
| PERIOD BIOL | B | 0.192 | 0.346 | 0.00050 | 0.03002 | 0.303 | 0.469 | 1.788 | 1.155 |
| PERSPECT BIOL MED | H&PS | 1.342 | 1.396 | 0.00217 | 0.19235 | 2.978 | 3.148 | 1.272 | 1.203 |
| PHILOS SCI | H&PS | 0.552 | 0.792 | 0.00219 | -- | 0.428 | 0.380 | 3.640 | 4.099 |
| PHILOS T R SOC B | B | 6.401 | 7.154 | 0.07729 | 0.49996 | 2.169 | 2.103 | 8.328 | 8.589 |
| PHYS LIFE REV | B | 7.208 | 5.241 | 0.00215 | -- | 2.381 | 1.816 | 8.543 | 11.201 |
| PHYS PERSPECT | H&PS | 0.214 | 0.262 | 0.00013 | -- | 0.302 | 0.319 | 2.000 | 1.893 |
| PHYS REV D | A&A | 4.558 | 4.027 | 0.30080 | 0.61496 | 4.068 | 3.760 | 3.162 | 3.421 |
| PLANET SPACE SCI | A&A | 2.224 | 2.128 | 0.01863 | 0.48178 | 2.896 | 2.981 | 2.167 | 2.105 |
| PLOS BIOL | B | 11.452 | 13.630 | 0.14959 | 1.19212 | 3.263 | 3.159 | 9.904 | 10.230 |
| PLOS ONE | B | 4.092 | 4.537 | 0.50162 | -- | 1.308 | 0.980 | 8.828 | 11.783 |
| PROG AEROSP SCI | EA | 3.000 | 3.554 | 0.00226 | 0.32586 | 0.859 | 0.692 | 9.856 | 12.234 |
| PUBL ASTRON SOC AUST | A&A | 2.259 | 2.370 | 0.00370 | 0.23862 | 5.819 | 6.049 | 1.096 | 1.054 |
| PUBL ASTRON SOC JPN | A&A | 2.438 | 3.108 | 0.01847 | 0.50170 | 3.962 | 4.406 | 1.737 | 1.562 |
| PUBL ASTRON SOC PAC | A&A | 3.582 | 2.997 | 0.01871 | 0.35906 | 5.132 | 5.227 | 1.970 | 1.934 |
| Q REV BIOL | B | 7.727 | 6.538 | 0.00320 | 1.00320 | 2.944 | 2.944 | 7.407 | 7.407 |
| RADIAT ENVIRON BIOPH | B | 1.696 | 1.755 | 0.00290 | 0.34765 | 1.755 | 1.764 | 2.727 | 2.713 |
| RADIAT RES | B | 2.684 | 2.844 | 0.01397 | 0.43866 | 1.864 | 1.712 | 4.063 | 4.424 |
| RADIO SCI | A&A | 1.075 | 1.124 | 0.00453 | 0.28768 | 1.561 | 1.736 | 1.943 | 1.747 |
| RES ASTRON ASTROPHYS | A&A | 1.320 | 1.325 | 0.00228 | -- | 3.880 | 4.643 | 0.960 | 0.802 |
| REV BIOL TROP | B | 0.459 | 0.544 | 0.00203 | 0.04986 | 0.646 | 0.732 | 2.005 | 1.770 |
| REV MEX ASTRON ASTR | A&A | 1.000 | 1.352 | 0.00158 | 0.16629 | 4.036 | 4.300 | 0.699 | 0.656 |
| REV MEX FIS E | H&PS | 0.111 | -- | 0.00011 | -- | 0.900 | 1.295 | 0.348 | 0.242 |
| RIV BIOL-BIOL FORUM | B | 0.613 | 0.455 | 0.00022 | 0.00432 | 2.643 | 2.643 | 0.655 | 0.655 |
| SCI CHINA LIFE SCI | B | 2.024 | 2.030 | 0.00092 | -- | 1.283 | 1.110 | 4.452 | 5.146 |
| SCI CHINA SER C | B | 1.610 | 1.148 | 0.00239 | 0.11358 | 1.353 | 1.353 | 3.358 | 3.358 |
| SCI CONTEXT | H&PS | 0.395 | 0.382 | 0.00044 | 0.06799 | 0.297 | 0.271 | 3.753 | 4.113 |
| SCI EDUC-NETHERLANDS | H&PS | 0.702 | -- | 0.00112 | -- | 0.919 | 1.099 | 2.156 | 1.803 |
| SCI ENG ETHICS | H&PS | 0.738 | 0.937 | 0.00112 | 0.13089 | 0.901 | 0.959 | 2.311 | 2.172 |
| SOC HIST MED | H&PS | 0.545 | 0.659 | 0.00090 | 0.16879 | 0.860 | 0.958 | 1.788 | 1.605 |
| SOC STUD SCI | H&PS | 1.500 | 2.286 | 0.00365 | 0.16842 | 0.449 | 0.290 | 9.428 | 14.597 |
| SOL PHYS | A&A | 2.776 | 2.880 | 0.02149 | 0.62204 | 3.677 | 3.895 | 2.131 | 2.011 |
| SOLAR SYST RES+ | A&A | 0.682 | 0.623 | 0.00106 | 0.08108 | 1.640 | 2.032 | 1.174 | 0.947 |
| SPACE SCI REV | A&A | 3.611 | 3.914 | 0.02961 | 0.72941 | 4.024 | 4.042 | 2.532 | 2.521 |
| SPACE WEATHER | A&A | 1.329 | 1.505 | 0.00227 | 0.29996 | 2.789 | 3.439 | 1.345 | 1.091 |
| STUD HIST PHILOS M P | H&PS | 0.641 | 0.622 | 0.00132 | 0.19796 | 1.172 | 1.356 | 1.543 | 1.334 |
| STUD HIST PHILOS SCI | H&PS | 0.513 | 0.677 | 0.00115 | 0.10470 | 0.381 | 0.323 | 3.800 | 4.482 |
| SYNTHESE | H&PS | 0.649 | 0.728 | 0.00309 | 0.09312 | 0.230 | 0.163 | 7.963 | 11.236 |
| T JPN SOC AERONAUT S | EA | 0.338 | 0.269 | 0.00041 | 0.07089 | 1.939 | 2.046 | 0.492 | 0.466 |
| TECHNOL CULT | H&PS | 0.321 | 0.422 | 0.00089 | 0.02158 | 0.131 | 0.029 | 6.915 | 31.237 |
| THEOR BIOSCI | B | 0.979 | 1.000 | 0.00069 | 0.23667 | 1.831 | 2.279 | 1.509 | 1.212 |
| TURK J BIOL | B | 0.876 | -- | 0.00062 | -- | 0.848 | 0.826 | 2.915 | 2.993 |
| ZH OBSHCH BIOL | B | 0.254 | 0.386 | 0.00033 | 0.05962 | 0.592 | 0.930 | 1.211 | 0.771 |

*Source: 2011 JCR, and Leydesdorff & Bornmann (2011); ES = Eigenfactor Score; FCIF = Fractionally Counted Impact Factor; CP = Citation Potential; TNIF = Topic Normalized Impact Factor; Self-cite = including self-citation; A&A = Astronomy & Astrophysics; B = Biology; EA = Engineering, Aerospace; H&PS = History & Philosophy of Science.*

Table 3: Pearson correlation coefficients

| *JCR Category* | *# Journals* | | *5-JIF* | *ES* | *FCIF* | *Self-cite* | *TNIF* |
|---|---|---|---|---|---|---|---|
| *Astronomy & Astrophysics* | *56* | *2-JIF* | 0.99*** | 0.22 | 0.94*** | 0.97*** | 0.98*** |
| | | *5-JIF* | | 0.15 | 0.94*** | 0.96*** | 0.97*** |
| | | *ES* | | | 0.17 | 0.20 | 0.21 |
| | | *FCIF* | | | | 0.95*** | 0.95*** |
| | | *Self-cite* | | | | | 1.00*** |
| *Biology* | *85* | *2-JIF* | 0.97*** | 0.36*** | 0.87*** | 0.85*** | 0.82*** |
| | | *5-JIF* | | 0.38*** | 0.89*** | 0.84*** | 0.80*** |
| | | *ES* | | | 0.51*** | 0.45*** | 0.49*** |
| | | *FCIF* | | | | 0.79*** | 0.73*** |
| | | *Self-cite* | | | | | 0.97*** |
| *Engineering, Aerospace* | *27* | *2-JIF* | 0.94*** | 0.27 | 0.41 | 0.84*** | 0.86*** |
| | | *5-JIF* | | 0.30 | 0.49* | 0.88*** | 0.91*** |
| | | *ES* | | | 0.20 | 0.26 | 0.27 |
| | | *FCIF* | | | | 0.37 | 0.41 |
| | | *Self-cite* | | | | | 0.96*** |
| *History & Philosophy of Science* | *56* | *2-JIF* | 0.96*** | 0.79*** | 0.62*** | 0.44*** | 0.31 |
| | | *5-JIF* | | 0.85*** | 0.64*** | 0.51*** | 0.33* |
| | | *ES* | | | 0.62*** | 0.58*** | 0.42** |
| | | *FCIF* | | | | 0.23 | 0.04 |
| | | *Self-cite* | | | | | 0.73*** |
| *Total* | *224* | *2-JIF* | 0.99*** | 0.32*** | 0.91*** | 0.64*** | 0.48*** |
| | | *5-JIF* | | 0.28*** | 0.91*** | 0.65*** | 0.47*** |
| | | *ES* | | | 0.31*** | 0.24*** | 0.19** |
| | | *FCIF* | | | | 0.58*** | 0.40*** |
| | | *Self-cite* | | | | | 0.85*** |

**** = 99% confidence level; ** = 95% confidence level; * = 90% confidence level; ES = Eigenfactor Score; FCIF = Fractionally Counted Impact Factor; TNIF = Topic Normalized Impact Factor; Self-cite = TNIF including self-citation.*

Table 4: Spearman rank correlation coefficients

| *JCR Category* | *# Journals* | | *5-JIF* | *ES* | *FCIF* | *Self-cite* | *TNIF* |
|---|---|---|---|---|---|---|---|
| *Astronomy & Astrophysics* | *56* | *2-JIF* | 0.98*** | 0.82*** | 0.87*** | 0.93*** | 0.95*** |
| | | *5-JIF* | | 0.82*** | 0.88*** | 0.88*** | 0.91*** |
| | | *ES* | | | 0.76*** | 0.80*** | 0.82*** |
| | | *FCIF* | | | | 0.82*** | 0.84*** |
| | | *Self-cite* | | | | | 0.99*** |
| *Biology* | *85* | *2-JIF* | 0.97*** | 0.82*** | 0.81*** | 0.83*** | 0.84*** |
| | | *5-JIF* | | 0.85*** | 0.82*** | 0.85*** | 0.85*** |
| | | *ES* | | | 0.71*** | 0.87*** | 0.86*** |
| | | *FCIF* | | | | 0.69*** | 0.68*** |
| | | *Self-cite* | | | | | 0.99*** |
| *Engineering, Aerospace* | *27* | *2-JIF* | 0.87*** | 0.75*** | 0.54** | 0.70*** | 0.84*** |
| | | *5-JIF* | | 0.77*** | 0.72*** | 0.69*** | 0.87*** |
| | | *ES* | | | 0.54** | 0.64*** | 0.77*** |
| | | *FCIF* | | | | 0.43 | 0.52** |
| | | *Self-cite* | | | | | 0.85*** |
| *History & Philosophy of Science* | *56* | *2-JIF* | 0.94*** | 0.66*** | 0.52*** | 0.35* | 0.57*** |
| | | *5-JIF* | | 0.70*** | 0.57*** | 0.34* | 0.48*** |
| | | *ES* | | | 0.49*** | 0.33* | 0.49*** |
| | | *FCIF* | | | | 0.03 | 0.22 |
| | | *Self-cite* | | | | | 0.75*** |
| *Total* | *224* | *2-JIF* | 0.98*** | 0.84*** | 0.65*** | 0.57*** | 0.65*** |
| | | *5-JIF* | | 0.85*** | 0.65*** | 0.60*** | 0.65*** |
| | | *ES* | | | 0.74*** | 0.54*** | 0.62*** |
| | | *FCIF* | | | | 0.39*** | 0.45*** |
| | | *Self-cite* | | | | | 0.91*** |

**** = 99% confidence level; ** = 95% confidence level; * = 90% confidence level; ES = Eigenfactor Score; FCIF = Fractionally Counted Impact Factor; TNIF = Topic Normalized Impact Factor; Self-cite = TNIF including self-citation.*

Table 5: Central-tendency and variability measures

| *JCR Category* | *Measures* | *2-JIF* | *5-JIF* | *ES* | *FCIF* | *Self-cite* | *TNIF* |
|---|---|---|---|---|---|---|---|
| *Astronomy & Astrophysics* | *Median* | 1.683 | 1.757 | 0.00430 | 0.31919 | 1.844 | 1.723 |
| | *Mean* | 3.070 | 3.180 | 0.03561 | 0.41331 | 2.144 | 2.112 |
| | *Sd* | 4.292 | 4.548 | 0.08311 | 0.39276 | 2.209 | 2.457 |
| *Biology* | *Median* | 1.540 | 1.719 | 0.00256 | 0.20534 | 2.915 | 2.993 |
| | *Mean* | 2.096 | 2.374 | 0.01595 | 0.26865 | 3.473 | 3.671 |
| | *Sd* | 2.115 | 2.375 | 0.05812 | 0.26257 | 2.640 | 3.086 |
| *Engineering, Aerospace* | *Median* | 0.549 | 0.654 | 0.00126 | 0.13113 | 1.886 | 1.507 |
| | *Mean* | 0.680 | 0.833 | 0.00283 | 0.14485 | 2.174 | 2.130 |
| | *Sd* | 0.605 | 0.734 | 0.00377 | 0.12627 | 1.868 | 2.390 |
| *History & Philosophy of Science* | *Median* | 0.442 | 0.553 | 0.00033 | 0.09312 | 2.134 | 1.810 |
| | *Mean* | 0.580 | 0.725 | 0.00077 | 0.11780 | 2.931 | 3.523 |
| | *Sd* | 0.603 | 0.636 | 0.00097 | 0.07779 | 2.408 | 5.274 |

*Sd = Standard deviation; ES = Eigenfactor Score; FCIF = Fractionally Counted Impact Factor; TNIF = Topic Normalized Impact Factor; Self-cite = TNIF including self-citation.*

Table 6: Central-tendency and variability measures for the aggregate data

| *Measures* | *2-JIF* | *5-JIF* | *ES* | *FCIF* | *Self-cite* | *TNIF* |
|---|---|---|---|---|---|---|
| *Median* | 1.000 | 1.159 | 0.00163 | 0.19203 | 2.111 | 1.922 |
| *Mean* | 1.790 | 1.998 | 0.01549 | 0.26395 | 2.849 | 3.059 |
| *Within-group variance ($Sd^2$)* | 7.325 | 8.124 | 0.00315 | 0.08509 | 5.995 | 13.128 |
| *Between-group variance ($Sd^2$)* | 1.432 | 1.441 | 0.00026 | 0.01826 | 0.412 | 0.730 |
| *Total reduction of the variance* | 5.893 | 6.683 | 0.00289 | 0.06683 | 5.583 | 12.398 |
| *Percentage reduction of the variance* | 80.5% | 82.3% | 91.9% | 78.5% | 93.1% | 94.4% |

*Sd = Standard deviation; ES = Eigenfactor Score; FCIF = Fractionally Counted Impact Factor; TNIF = Topic Normalized Impact Factor; Self-cite = TNIF including self-citation; Within-group = within the set of all journals (224); Between-group = between the JCR categories.*